\documentclass[a4paper]{amsart}
\date{January 15, 2021}
\def\arsinh{\mathfrak{Arsin}}

\def\rz{\mathbb{R}}
\def\cE{\mathcal{E}}

\def\rd{\mathrm{d}}
\def\cE{\mathcal{E}}
\def\cH{\mathcal{H}}
\def\cT{\mathcal{T}}
\def\cTg{{{\mathcal{T}}_>}}
\def\gR{\mathfrak{R}}
\def\gZ{\mathfrak{Z}}
\def\D{\mathcal{D}}
\def\tf{T^\mathrm{TF}}
\def\etf{{e^\mathrm{TF}}}
\def\ETF{{\cE_Z^\mathrm{TF}}}
\def\TF{\mathcal{T}^\mathrm{TF}}
\def\TFW{\mathcal{E}_{c,Z}^\mathrm{TFWD}}
\def\V{\mathcal{V}}
\def\W{\mathcal{T}^\mathrm{W}}
\def\x{X}
\def\X{\mathcal{X}}
\def\gtf{{\gamma_\mathrm{TF}}}
\def\a{{A_{s,Z}}}
\def\i{{I_{s,Z}}}
\newcommand{\dbar}{{\mathchar'26\mkern-12mu\mathrm{d}}}

\newtheorem{lemma}{Lemma}
\newtheorem{satz}{Theorem}

\title[Statistical Theory of Heavy Atoms]{A Statistical Theory of
  Heavy Atoms:\\ Asymptotic Behavior of the Energy and Stability of Matter}
\begin{document}
\author[H. Siedentop]{Heinz Siedentop} \address{Mathematisches
  Institut\\ Ludwig-Maximilans Universit\"at M\"unchen,
  Theresienstra\ss e 39\\ 80333 M\"unchen\\ Germany\\ and Munich Center for
  Quantum Science and Technology (MCQST)\\ Schellingstr. 4\\ 80799
  M\"unchen, Germany} \email{h.s@lmu.de}

\dedicatory{Dedicated to Ari Laptev on the occasion of his
  septuagesimal birthday.\\ His ideas in analysis have inspired many.}
\begin{abstract}
  We give the asymptotic behavior of the ground state energy of
  Engel's and Dreizler's relativistic Thomas-Fermi-Weizs\"acker-Dirac
  functional for heavy atoms for fixed ratio of the atomic number and
  the velocity of light. Using a variation of the lower bound, we show
  stability of matter.
\end{abstract}
\maketitle
\section{Introduction \label{Einleitung}}

Heavy atoms require a relativistic description because of the
extremely fast moving inner electrons. However, a statistical theory
of the atom in the the spirit of Thomas \cite{Thomas1927} and Fermi
\cite{Fermi1927,Fermi1928} yields a functional which is unbounded from
below because the semi-classical relativistic Fermi energy is too weak
to prevent mass from collapsing into the nucleus. (See Gombas \cite[\S
14]{Gombas1949} and \cite[Chapter III, Section
16.]{Gombas1956}. Gombas also suggested that Weizs\"acker's
(non-relativistic) inhomogeneity correction would solve this
problem. Tomishia \cite{Tomishima1969} carried this suggestion
through.) Because of the same reason the relativistic generalization
of the Lieb-Thirring inequality by Daubechis is not directly
applicable to Chandrasekhar operators with Coulomb potentials but
requires a separate treatment of the singularity. Frank and Ekholm
\cite{FrankEkholm2006} found a way circumventing this problem treating
critical potentials of Schr\"odinger operators; later Frank et
al. \cite{Franketal2007S} accomplished the same for Chandrasekhar
operators with Coulomb potentials. It amounts to a Thomas-Fermi
functional with a potential whose critical singularity has been
extracted. However, there is a drawback, namely the Thomas-Fermi
constant of this functional is smaller than the classical one, i.e.,
we cannot expect asymptotically correct results.

Here we discuss an alternative relativistic density functional which
can handle Coulomb potentials of arbitrary strength: Engel and
Dreizler \cite{EngelDreizler1987} derived a functional $\TFW$ of the
electron density $\rho$ from quantum electrodynamics (in contrast to
Gombas' ad hoc procedure of adding the non-relativistic Weizs\"acker
correction). It is -- in a certain sense -- a generalization of the
non-relativistic Thomas-Fermi-Weizs\"acker-Dirac functional to the
relativistic setting, a feature that it shares with the functional
investigated by Lieb et \cite{Liebetal1996}. However, it does not
suffer from the problem that it becomes unbounded from below for heavy
atoms. We will show here, that it has -- unlike the functional which
can be obtained from \cite{Franketal2007S} -- the same asymptotic
behavior as the atomic quantum energy. The price to pay is the absence
of a known inequality relating it to the full quantum problem. One
could speculate that it might be an upper bound on the ground state
energy. The way we prove the upper bound of the asymptotics might
nourishes such thoughts. However, even in the non-relativistic context
this is open despite numerical evidence and claims to the contrary,
e.g., by March and Young \cite{MarchYoung1958}: the arguments given
contain a gap. In other words, such a claim would have -- even in the
non-relativistic context -- at best the status of a conjecture.

Engel's and Dreizler's functional is the relativistic TF functional
(see Chandrasekhar \cite{Chandrasekhar1931} [in the ultrarelativistic
limit] and Gombas \cite[\S 14]{Gombas1949} for the general case) with
an inhomogeneity and exchange correction different from the
non-relativistic terms but with an integrand tending pointwise to
their non-relativistic analogue for large velocity of light $c$. In
Hartree units it reads for atoms of atomic number $Z$ and electron
density $\rho$
\begin{equation}
\label{TFWD}
\TFW(\rho)
:= \W(\rho)+\TF(\rho)-\X(\rho) +\V(\rho). 
\end{equation}
The first summand on the right is an inhomogeneity correction of the
kinetic energy generalizing the Weizs\"acker correction.  Using the
Fermi momentum $p(x):= (3\pi^2 \rho(x))^{1/3}$ it is
\begin{equation}
  \label{W}
  \W(\rho):=
  \int_{\rz^3}\rd x{3\lambda\over8\pi^2}(\nabla p(x))^2c\,f(p(x)/c)^2
\end{equation}
with $ f(t)^2:=t(t^2+1)^{-\frac12}+2t^2(t^2+1)^{-1}\arsinh(t)$ where
$\arsinh$ is the inverse function of the hyperbolic sine. The
parameter $\lambda\in\rz_+$ is given by the gradient expansion as
$1/9$ but is in the non-relativistic analogue sometimes taken as an
adjustable parameter (Weizs\"acker \cite{Weizsacker1935}, Yonei and
Tomishima \cite{YoneiTomishima1965}, Lieb
\cite{LiebLiberman1982,Lieb1982A}).

The second summand is the relativistic generalization of the
Thomas-Fermi kinetic energy, namely
\begin{equation}
  \label{TF}
  \TF(\rho):=\int_{\rz^3}\rd x{c^5\over8\pi^2} \tf(\tfrac{p(x)}c) 
\end{equation}
with
$\tf(t):=t(t^2+1)^{3/2}+t^3(t^2+1)^{1/2}-\arsinh(t)-{8\over3}t^3$.

The third summand is a relativistic generalization of the exchange
energy. It is
\begin{equation}
  \label{X}
  \X(\rho):= \int_{\rz^3}\rd x{c^4\over8\pi^3} \x(\tfrac{p(x)}c)
\end{equation}
with $\x(t):= 2t^4-3[t(t^2+1)^\frac12-\arsinh(t)]^2$.

Eventually, the last summand is the potential energy, namely the sum
of the electron-nucleus and the electron-electron interaction. It is
\begin{equation}
  \label{D}
  \V(\rho):= -Z\int_{\rz^3}\rd x\rho(x)|x|^{-1}+\underbrace{\tfrac12\int_{\rz^3}\rd x \int_{\rz^3}\rd y\rho(x)\rho(y)|x-y|^{-1}}_{=:\D[\rho]}.
\end{equation}
Using $F(t):=\int_0^t\rd s f(s)$, the functional $\TFW$ is naturally
defined on
\begin{equation}
  \label{domain}
  P:=\{\rho\in L^\frac43(\rz^3)|\rho\geq0,\ \D[\rho]<\infty, F\circ p\in D^1(\rz^3)\}
\end{equation}
and bounded from below \cite{ChenSiedentop2020} for all $c$ and $Z$.
In fact Chen et al. \cite{Chenetal2020} obtained a Thomas-Fermi type
lower bound for fixed ratio $\kappa:=Z/c$. Unfortunately its
Thomas-Fermi constant $\gamma_e$ is less than
$\gtf:=(3\pi^2)^\frac23$, the correct physical value.

For comparison we need the non-relativistic Thomas-Fermi functional
\begin{equation}
  \label{tf}
  \ETF(\rho):= \frac3{10}\gtf \int_{\rz^3}\rd x\rho(x)^\frac53 +\V(\rho)
\end{equation}
defined on $I:=\{\rho\in L^\frac53(\rz^3)|\rho\geq0,\
\D[\rho]<\infty\}$. The functional is bounded from below (Simon
\cite{Simon1979}) and its infimum fulfills the scaling relation
\begin{equation}
  \label{tfenergie}
  E^\mathrm{TF}(Z):= \inf\ETF(I) = -\etf Z^\frac73
\end{equation}
where $\etf=-E^\mathrm{TF}(1)$ (Gombas \cite{Gombas1949}, Lieb and
Simon \cite{LiebSimon1977}). There exists a unique minimizer of $\ETF$
which we denote by $\sigma$.

Our first result is
\begin{satz}
  Assume $\kappa:=Z/c\in\rz_+$ fixed. Then, as $Z\to\infty$,
  \begin{equation}
    \label{resultat}
    \inf\TFW(P)= - e^\mathrm{TF}Z^\frac73+O(Z^2).
  \end{equation}
\end{satz}
Our second result is the stability of the second kind of the functional
which we address in Section \ref{Molekuele}.

From a mathematical perspective it might come as a surprise that
Engel's and Dreizler's density functional -- derived by purely formal
methods from a quantum theory which is still lacking a full
mathematical understanding -- yields a fundamental feature like the
ground state energy of heavy atoms to leading order quantitatively
correct, in full agreement with the $N$-particle descriptions of heavy
atoms like the Chandrasekhar Hamiltonian and no-pair Hamiltonians (S\o
rensen \cite{Sorensen2005}, \cite{CassanasSiedentop2006}, Solovej
\cite{Solovejetal2008}, Frank et
al. \cite{Franketal2008,Franketal2009},
\cite{HandrekSiedentop2013}). It remains to be seen whether this is
also true for other quantities like the density or whether the
functional can be used as a tool to investigate relativistic many
particle systems -- like Thomas-Fermi theory in non-relativistic many
body quantum mechanics -- or, whether it even can shed light on a
deeper understanding of quantum electrodynamics.

\section{Bounds on the Energy \label{ua}}

\subsection{Upper Bound on the Energy}
We begin with an innocent lemma.
\begin{lemma}
  \label{0}
  Assume $\rho:\rz^3\to\rz_+$ such that $p:= (3\pi^2\rho)^\frac13$ has
  partial derivatives with respect to all variables at
  $x\in\rz^3$. Then
  \begin{equation}
    \label{os1}
    {3\over8\pi^2 }|\nabla p (x)|^2cf(p(x)/c)
      \leq  |(\nabla\sqrt\rho)(x)|^2.
    \end{equation}
\end{lemma}
Thus every nonnegative $\rho$ with $\nabla\sqrt\rho\in L^2(\rz^3)$ fulfills
\begin{equation}
  \label{os2}
  \W(\rho)  \leq\lambda\int_{\rz^3}|\nabla\sqrt\rho|^2.
\end{equation}
\begin{proof}
  We set $\psi:=\sqrt\rho$ and compute
  \begin{equation}
    \label{os2c}
    \begin{split}
      &{3\over8\pi^2} |\nabla p (x)|^2cf(p(x)/c)\\
      =& {3^2\over8}
      |\nabla\sqrt[3]\rho(x)|^2\left({\psi^\frac23(x)\over\sqrt{1+(p(x)/c)^2}}
        +2{\psi^\frac23(x)(p(x)/c)\arsinh(p(x)/c)\over 1 +(p(x)/c)^2}\right)\\
      \leq& \tfrac12 |\nabla\psi(x)|^2\max\left\{{\sqrt{1+t^2}+2t\arsinh(t)\over1+t^2}|t\in\rz_+\right\}
      \leq |\nabla\psi(x)|^2 .
    \end{split}
  \end{equation}
\end{proof}
Of course the illuminati are hardly impressed by \eqref{os2}, since
dominating relativistic energies by non-relativistic ones is common
place for them. Presumably not even use of the numerically correct
value 1.658290113 of the maximum in the proof instead of the estimate
2 would change that.

Now we turn to the upper bound on the left side of \eqref{resultat}.
It will be practical to use the non-relativistic
Thomas-Fermi-Weizs\"acker functional
\begin{equation}
  \label{tfw}
  \cE^\mathrm{nrTFW}_Z(\rho):= \frac \beta2\int_{\rz^3}|\nabla\sqrt\rho|^2 + \cE^\mathrm{TF}_Z(\rho)
\end{equation}
where $\beta\in\rz_+$. It is defined on
$J:=\{\rho\in L^\frac53(\rz^3)|\rho\geq0,\ \sqrt\rho\in D^1(\rz^3),\
\D[\rho]<\infty\}$, has a unique minimizer $\rho_W$ with
$\int_{\rz^3}\rho_W\leq Z+C$, and
$\int_{\rz^3}\rho_W^\frac53=O(Z^\frac73)$ (Benguria
\cite{Benguria1979}, Benguria et al. \cite{Benguriaetal1981}, Benguria
and Lieb \cite{BenguriaLieb1985}). Moreover,
\begin{equation}
  \label{tfwenergie}
  E^\mathrm{nrTFW}(Z)=\cE^\mathrm{nrTFW}_Z(\rho_W) =E^\mathrm{TF}(Z)+ D_\beta Z^2+o(Z^2)
\end{equation}
for some $\beta$-dependent constant $D_\beta\in\rz_+$ (Lieb and Liberman
\cite{LiebLiberman1982} and Lieb \cite[Formula (1.6)]{Lieb1982A}).

In the following we pick $\beta=2$ and use the minimizer $\rho_W$ of the
non-relativistic Thomas-Fermi-Weizs\"acker functional as a test
function.

We estimate the exchange term first. Since $-X(t)\leq t^4$, we get
\begin{equation}
  \label{eX}
    -\X(\rho_W)\leq {(3\pi^2)^\frac43\over 8\pi^3} \int_{\rz^3} \rho_w^\frac43
    \leq C\sqrt{\int_{\rz^3}\rho_W^\frac53 \int_{\rz^3}\rho_W}=O(Z^\frac53).
\end{equation}
Thus, since $\tf(t)\leq \frac35t^5$,
\begin{equation}
  \label{ergeb}
  \inf\TFW(P)\leq \TFW(\rho_W) \leq \cE^\mathrm{nrTFW}_Z(\rho_W)+ O(Z^\frac53)= E^\mathrm{TF}(Z) +O(Z^2)
\end{equation}
which concludes the proof of the upper bound.

\subsection{Lower Bound on the Energy}

We set $\dbar \xi:=\rd \xi/h^3=\rd\xi/(2\pi)^3$. (Note that the
rationalized Planck constant $\hbar$ equals one in Hartree units and,
therefore, $h=2\pi$.)

We introduce the notation $(a)_-:=\min\{0,a\}$ and write
$\varphi_\sigma:=Z/|\cdot|-\sigma*|\cdot|^{-1}$ for the
Thomas-Fermi potential of the minimizer $\sigma$. We start again with
a little Lemma.
\begin{lemma}
  \label{l}
  Assume $\kappa=Z/c$ fixed. Then, as $Z\to \infty$,
  \begin{equation*}
    \begin{split}
      \int\limits_{|x|>\frac1Z}\rd x \int\limits_{\rz^3}\dbar \xi(\frac{\xi^2}2-\varphi_\sigma(x))_- -\int\limits_{|x|>\frac1Z}\rd x \int\limits_{\rz^3}\dbar \xi(\sqrt{c^2\xi^2+c^4}-c^2-\varphi_\sigma(x))_- =O(Z^2).
      \end{split}
  \end{equation*}
\end{lemma}
Again, it does not come as a surprise to the physicist that
relativistic and non-relativistic theory give the same result up to
errors, if the innermost electrons, i.e., in particular the fast
moving, are disregarded.
\begin{proof}
  Since $\xi^2/2\geq \sqrt{c^2\xi^2+c^4}-c^2$, the left side of the
  claimed inequality cannot be negative. Thus, we merely need an upper
  bound:
  \begin{equation}
    \begin{split}
    &\int\limits_{|x|>\frac1Z} \rd x \int\limits_{\rz^3}\dbar \xi(\frac{\xi^2}2-\varphi_\sigma(x))_- -\int\limits_{|x|>\frac1Z}\rd x \int\limits_{\rz^3}\dbar \xi\left(\sqrt{c^2\xi^2+c^4}-c^2-\varphi_\sigma(x)\right)_-\\
    \leq &\int_{|x|>\frac1Z}\rd x \left\{c^5\int_{{\xi^2\over2}<{\varphi_\sigma(x)\over c^2}}\dbar \xi[\tfrac12{\xi^2} -(\sqrt{\xi^2+1}-1)]\right.\\
   &\left. - \int_{\sqrt{c^2\xi^2+c^4}-c^2<\varphi_\sigma(x)\leq\xi^2/2}\dbar \xi\left(\sqrt{c^2\xi^2+c^4}-c^2-\varphi_\sigma(x)\right)\right\}\\
    \leq &c^5\int_{|x|>\frac1Z}\rd x \left(\int\limits_{{\xi^2\over2}<{\varphi_\sigma(x)\over c^2}}\dbar \xi+\int\limits_{\sqrt{\xi^2+1}-1<{\varphi_\sigma(x)\over c^2}\leq{\xi^2\over2}}\dbar \xi \right)[\tfrac12{\xi^2} -(\sqrt{\xi^2+1}-1)]\\
    \leq&{c^5\over8}\int_{|x|>\frac1Z}\rd x \left(\int_{{\xi^2\over2}<{\varphi_\sigma(x)\over c^2}}\dbar \xi +\int_{\sqrt{\xi^2+1}-1<{\varphi_\sigma(x)\over c^2}\leq{\xi^2\over 2}}\dbar \xi\right) |\xi|^4\\
    \leq&{c^5\over8} \int_{|x|>{1\over Z}} \rd x \int_{\sqrt{\xi^2+1}-1<{\varphi_\sigma(x)\over c^2}}\dbar \xi |\xi|^4\leq{c^2\over8\kappa^3} \int_{|x|>1} \rd x \int_{\sqrt{\xi^2+1}-1<{\kappa^2\over |x|}}\dbar \xi |\xi|^4
    \end{split}
  \end{equation}
  where we used $\varphi_\sigma(x)\leq Z/|x|$ in the last
  inequality. Moreover, the resulting last integral obviously exists and is
  independent of $Z$. Thus the left side of the claimed inequality is
  bounded from above by a constant depending only on $\kappa$ times
  $Z^2$ quod erat demonstrandum.
\end{proof}

We turn to the lower bound on the left side of \eqref{resultat} and
follow initially \cite{Chenetal2020}. In fact, apart from minor
modifications, we copy the high density part and focus on the low
density part. We pick any $\rho\in P$ and address the parts of the
energy separately.

\subsubsection{The Weizs\"acker Energy}
Since $F(t)\geq t\sqrt{\arsinh(t)}/2$ (see \cite[Formula
(90)]{ChenSiedentop2020}), Hardy's inequality gives the lower bound
\begin{equation}
  \label{usW}
  \W(\rho)
  \geq {3\lambda c\over2^7\pi^2} \int_{\rz^3}\rd x
  {p(x)^2\arsinh(\tfrac{p(x)}c)\over|x|^2}={3^\frac53\lambda c\over2^7\pi^\frac23} \underbrace{\int_{\rz^3}\rd x
  {\rho(x)^\frac23\arsinh(\tfrac{p(x)}c)\over|x|^2}}_{=:\cH(\rho)}.
\end{equation}

\subsubsection{The Potential Energy}
Since $\sigma$ is positive, we have $\varphi_\sigma(x)\leq
Z/|x|$. Then
\begin{equation}
  \label{ww}
  \V(\rho)= -\int_{\rz^3}\rd x \varphi_\sigma(x)\rho(x) -2\D(\sigma,\rho)+\D[\rho] 
  \geq -\int_{\rz^3}\rd x \varphi_\sigma(x)\rho(x) - \D[\sigma].
\end{equation}
Splitting the integrals at $s$, using \eqref{ww}, and Schwarz's
inequality yields
\begin{equation}
  \label{P}
  \begin{split}
    \V(\rho)
    \geq& -\int_{p(x)/c<s}\rd x\varphi_\sigma(x)\rho(x)\\
    &-Z\int_{p(x)/c\geq s}\rd x {\rho(x)^\frac13\over|x|}
    \arsinh(\tfrac{p(x)}c)^\frac12
    {\rho(x)^\frac23\over \arsinh(\tfrac{p(x)}c)^\frac12}-\D[\sigma]\\
    \geq&
    -{Z\over\arsinh(s)^\frac12}\cH(\rho)^\frac12\cTg(\rho)^\frac12
    -\int_{p(x)/c<s}\rd x\varphi_\sigma(x)\rho(x) -\D[\sigma]
    \end{split}
\end{equation}
with $\cTg(\rho):= \int_{p(x)/c>s}\rd x \rho(x)^\frac43 $.

\subsubsection{The Thomas-Fermi Term}
First, we note that
\begin{equation}
  \label{t4}
  \rz_+\to\rz_+,\ t\mapsto\tf(t)/t^4
\end{equation}
is monotone increasing from $0$ to $2$. Thus
\begin{equation}
  \label{tfl}
  \begin{split}
    &\TF(\rho)=\int_{p(x)/c<s}\rd x\frac{c^5}{8\pi^2}\tf(\tfrac{p(x)}c)+ \int_{p(x)/c\geq s}\rd x \frac{c^5}{8\pi^2}\tf(\tfrac{p(x)}c)\\
    \geq&\int_{p(x)/c<s}\rd x\frac{c^5}{8\pi^2}\tf(\tfrac{p(x)}c)+ \int_{p(x)/c\geq s}\rd x \frac{\tf(s)}{s^4}\tfrac38(3\pi^2)^\frac13c\rho(x)^\frac43\\
    =&\int_{p(x)/c<s}\rd x\frac{c^5}{8\pi^2}\tf(\tfrac{p(x)}c)+\frac38\frac{\tf(s)}{s^4}\gtf^\frac12c\cTg(\rho).
  \end{split}
\end{equation}

\subsubsection{Exchange Energy}
Since $X$ is bounded from above and $X(t)=O(t^4)$ at $t=0$, we have
that for every $\alpha\in[0,4]$ there is an $\eta_0$ such that
$X(t)\leq \eta_0 t^\alpha$. We pick $\alpha=3$ in which case
$\xi_0\approx 1.15$. Thus, with $\eta:=\eta_0/(4\pi)\approx 0.0914$,
we have
\begin{equation}
  \label{xl}
  \X(\rho)\leq {c\eta_0\over4\pi}N = \eta c N.
\end{equation}

\subsubsection{The Total Energy\label{ax}}
Adding everything up yields
\begin{equation}
  \label{hardyww90}
  \begin{split}
    \TFW(\rho) \geq &{3^\frac53\lambda c\over2^7\pi^\frac23} \cH(\rho)
    +\frac38\frac{\tf(s)}{s^4}\gtf^\frac12c\cTg(\rho)
    -{Z\over \arsinh(s)^\frac12}\cH(\rho)^\frac12\cTg(\rho)^\frac12\\
    & + \int_{\frac{p(x)}c<s}\rd x\frac{c^5}{8\pi^2}\tf(\tfrac{p(x)}c)
-\varphi_\sigma(x)\rho(x)) -\D[\sigma] -\xi c N.
  \end{split}
\end{equation}
We pick $s\in\rz_+$ such that the sum of the first three summands on
the right of \eqref{hardyww90} is a complete square, i.e., fulfilling
\begin{equation}
  \label{vollquad}
  \sqrt{{3^\frac53\over2^7\pi^\frac23}{3\tf(s)(3\pi^2)^\frac13\over8s^4}}
  ={Z\over c\sqrt\lambda}{1\over2\arsinh(s)^\frac12}.
\end{equation}
The solution is uniquely determined, because of \eqref{t4} (and the
line below) and $\arsinh(s)$ is also monotone increasing from $0$ to
$\infty$. Call the corresponding $s$ $s_0$. Obviously, $s_0$ does not
depend on $c$ and $Z$ independently but only on the ratio
$\kappa:=Z/c$ and is strictly monotone increasing from $0$ to
$\infty$.

We set
\begin{equation}
  \begin{split}
    \i:=&\{x\in\rz^3|p(x)/c<s,\ |x|<1/Z\},\\
    \a:=&\{x\in\rz^3|p(x)/c<s,\ |x|\geq1/Z\}.
  \end{split}
\end{equation}
Then
\begin{equation}
  \label{resultata}
  \TFW(\rho)\geq I+A  -\D[\sigma] - \xi c N
\end{equation}
with
\begin{align}   
  I:=&\int_\i\rd x\left(\frac{c^5}{8\pi^2}\tf(\tfrac{p(x)}c)
       -\varphi_\sigma(x)\rho(x)\right)\\
  A:=&\int_\a\rd x\left(\frac{c^5}{8\pi^2}\tf(\tfrac{p(x)}c)
       -\varphi_\sigma(x)\rho(x)\right)
\end{align}
We estimate $I$ from below by dropping the TF-term, using $\varphi_\sigma(x)\leq Z/|x|$, and observing that $x\in\i$. We get
\begin{equation}
  I\geq -C_\kappa Zc^3\int_0^{Z^{-1}}\rd r r =-C_\kappa Z^2
\end{equation}
where $C_\kappa$ is a generic constant depending on $\kappa$ only. In
other words, we can pull the Coulomb tooth paying a negligible price.

Next we estimate $A$ from below by keeping $\a$ fixed and minimizing
the integrand at each point $x\in\a$ by varying the values
$p(x)\in \rz_+$. We get
\begin{equation}
  \label{Au}
  \begin{split}
    A\geq &  2\int_\a\rd x \int_{\rz^3}\dbar \xi\left(\sqrt{c^2\xi^2+c^4}-c^2-\varphi_\sigma(x)\right)_-\\
    \geq & 2\int_{|x|\geq1/Z}\rd x \int_{\rz^3}\dbar \xi\left(\sqrt{c^2\xi^2+c^4}-c^2-\varphi_\sigma(x)\right)_-.
  \end{split}
\end{equation}
Although at first glance the first inequality might seem abrupt, it is
easily checked that the Thomas-Fermi functional (relativistic or
non-relativistic, restricted to some region in space $M$) with kinetic
energy $T(\xi)$ and external potential $\varphi$ is merely the
marginal functional (integrating out the momentum variable $\xi$) of
the phase space variational principle
\begin{equation}
  \label{phase}
  \cE_\Gamma(\gamma): = \int_M\rd x \int_{\rz^3}\dbar \xi (T(\xi)-\varphi(x))\gamma(x,\xi)
\end{equation}
with $\gamma(M,\rz^3)\subset[0,2]$ and the choice
$\gamma(x,\xi):= \chi_{\{(x,\xi)\in M\times \rz^3||\xi|<p(x)\}}$ for
given Fermi momentum $p$. Eventually, since $\gamma$ has only values
between $0$ and $2$, \eqref{phase} is obviously minimized by the
characteristic function of the support of the negative part of
$T(\xi)-\varphi(x)$ times 2.

Metaphorically speaking the second inequality of \eqref{Au} states,
that we can bound the energy from below by a toothless relativistic
Thomas-Fermi energy.

Now, by Lemma \ref{l} this equals the corresponding non-relativistic
expression up to errors of order $O(Z^2)$, i.e., we have
\begin{equation}
  \label{us}
  \begin{split}
    \TFW(\rho)\geq &2\int_{|x|\geq1/Z}\rd x \int_{\rz^3}\dbar \xi\left(\tfrac12\xi^2-\varphi_\sigma(x)\right)_--\D[\sigma]-C_\kappa Z^2\\
    \geq & 2\int_{\rz^3}\rd x \int_{\rz^3}\dbar \xi\left(\tfrac12\xi^2-\varphi_\sigma(x)\right)_--\D[\sigma] -C_\kappa Z^2\\
    = &\cE_Z^\mathrm{TF}(\sigma) -C_\kappa Z^2= -\etf Z^{7/3}-C_\kappa Z^2
    \end{split}
\end{equation}
which concludes the proof of the desired lower bound and therefore
also the proof of \eqref{resultat}.

\section{Stability of Matter\label{Molekuele}}

For several atoms the potential term $\V$ in \eqref{D} is
replaced by
\begin{equation}
  \label{Dn}
  \V(\rho):=-\sum_{k=1}^K\int_{\rz^3}\rd x {Z_k\rho(x)\over|x-R_k|}+\D[\rho] + \sum_{1\leq k<l\leq K}{Z_kZ_l\over|R_k-R_l|}
\end{equation}
with pairwise different positions of the nuclei
$\gR:=(R_1,...,R_K)\in\rz^{3K}$, and atomic numbers
$\gZ:=(Z_1,...,Z_K)\in\rz^K_+$. The first term is the attraction
potential between the electrons and the nuclei, the second term, the
electron-electron interaction, is unchanged, and the third is the
repulsion between the nuclei. We write $\cE_{\gR,\gZ}^\mathrm{TFWD}$
for the otherwise unchanged Engel-Dreizler functional. We are
interested in finding a lower bound to $\cE_{\gR,\gZ}^\mathrm{TFWD}$
which is uniform in the density $\rho$ as long as the constraint
$\int\rho= N$ is respected, is uniform in $R_1,...,R_K$ and linear in
$N$ and $K$, i.e., we wish to show that the energy per particle is
bounded from below by the same constant irrespective of the electron
density and the positions of the nuclei. This property is also known
as stability of the second kind, thermodynamic stability, and
stability of matter.

Benguria et al. \cite{Benguriaetal2008} showed stability of matter for
the massless version of this functional without the regularizing
$\arsinh$-term in $\W$ and without the exchange correction.

Initially we restrict to the case of equal atomic numbers $Z$. We
begin by extracting all Coulomb teeth which is conveniently done using
\cite[Formula (5.2)]{LiebYau1988}: given disjoint balls $B_1,...,B_K$
with centers at $R_1,...,R_K$ and radii $D_1,...,D_K$ we have
\begin{equation}
  \label{liebyau}
  -\Delta\geq \sum_{\kappa=1}^K\left({1\over4|\cdot-R_\kappa|^2}
    -D_k^{-2}Y(\tfrac{|\cdot-R_\kappa|}{D_\kappa})\right)_+\chi_{B_\kappa}
\end{equation}
with $Y(r)=1+r^2/4$. We also set
$H_k(x):=2\sqrt{Y(|x-R_k|/D_K)}/D_k$. We pick the radii maximal,
namely $D_k$ as half the distance of the $k$-th nucleus to its nearest
neighbor.

We also use Lieb and Yau's electrostatic inequality \cite[Formula
(4.5)]{LiebYau1988}: we write $\Gamma_1,...,\Gamma_K$ for the Voronoi
cells of $R_1,...,R_K$ and set
\begin{equation}
  \label{Phi}
  \Phi(x) :=\sum_{l=1}^K\Gamma_l(x)\sum_{k=1,k\neq l}^K{Z\over|x-R_k|},
\end{equation}
i.e., the nuclear potential at point $x$ of all nuclei except the one
from the cell in which $x$ lies. With this notation the electrostatic
inequality reads
\begin{equation}
  \label{electrostat}
  \tfrac12 D[\nu]-Z\int_{\rz^3}\Phi(x)\rd\nu(x)
  +Z^2\sum_{1\leq k<l\leq K}{1\over|R_k-R_l|}
  \geq {Z^2\over8}\sum_{k=1}^K{1\over D_k}
\end{equation}
for any bounded measure and $Z\in\rz_+$.

With these tools we modify the atomic lower bounds term by term.

\subsection{The Weizs\"acker Energy}
Instead of \eqref{usW} we get
\begin{equation}
  \label{usWn}
  \W(\rho)
  \geq {3^\frac53\lambda c\over2^7\pi^\frac23} \sum_{k=1}^K
  \underbrace{\int_{B_k}\rd x\rho(x)^\frac23\arsinh(\tfrac{p(x)}c)\left(|x-R_k|^{-2}
    -H_k(x)^2\right)_+}_{=:\cH_k(\rho)}
\end{equation}
where we use \eqref{liebyau} and $F(t)\geq t\sqrt{\arsinh(s)}/2$ (see
\cite[Formula (90)]{ChenSiedentop2020}).

\subsection{The Potential Energy}
Using \eqref{electrostat} and $\sqrt{(a^2-b^2)_+}\geq a-b$ for
$a,b\in\rz_+$ we get
\begin{equation}
  \label{wm}
  \begin{split}
    &\V(\rho) \geq -\sum_{k=1}^K\left(\int_{B_K}\rd x +\int_{\Gamma_k\setminus B_k}\rd x\right){Z\rho(x)\over|x-R_k|}+ {Z^2\over8}\sum_{k=1}^K{1\over D_k}\\
     \geq&-\sum_{k=1}^KZ\left\{\int_{B_k,\ p(x)/c\geq s}\rd x
      \rho(x)\left[\sqrt{\left({1\over|x-R_k|^2}-H_k(x)^2\right)_+} +H_k(x)\right]\right.\\
    &+\left.\int_{B_k,\ p(x)/c\leq s}\rd x {\rho(x)\over|x-R_k|}+\int_{\Gamma_k\setminus B_k}{\rho(x)\over|x-R_k|}-\frac Z8{1\over D_k}\right\}\\
      \geq& -\sum_{k=1}^K\left({Z\over\sqrt{\arsinh(s)}}\sqrt{\cH_{R_k}(\rho)\cT_{R_k}(\rho)}+Z\int_{B_k,\ p(x)/c\geq s}\rd xH_k(x)\rho(x)\right.\\
       &+\left.\int_{B_k,\ p(x)/c\leq s}\rd x {Z\rho(x)\over|x-R_k|}+\int_{\Gamma_k\setminus B_k}\rd x{Z\rho(x)\over|x-R_k|}\right)+{Z^2\over 8}\sum_{k=1}^K{1\over D_k}
  \end{split}
\end{equation}
with $\cT_{R_k}(\rho):= \int_{B_k,\ p(x)/c\geq s}\rd x\rho(x)^\frac43$.

\subsection{The Combined Thomas-Fermi and Exchange Terms}
We use $\tf(t)\geq 2t^4-8t^3/3$, \eqref{xl}, and set
$\delta:=(3\pi^2)^\frac13$. This yields
\begin{equation}
  \label{tfls}
  \TF(\rho)-\X(\rho)=\int\limits_{\rz^3}\rd x\frac{c^5}{8\pi^2}\tf(\tfrac{p(x)}c)-\eta cN \geq \tfrac34\delta c\int\limits_{\rz^3}\rd x \rho(x)^\frac43-C_cN,
\end{equation}

\subsection{The Total Energy}
Adding again all up yields
\begin{align}
  \label{1}
       &\TFW(\rho)\\
  \label{2}
  \geq& \sum_{k=1}^Kc\left[{3^\frac53\lambda
          \over2^7\pi^\frac23} \cH_{R_k}(\rho)
        +\tfrac38\delta\cT_{R_k}(\rho)
        -{\kappa\over \sqrt{\arsinh(s)}}\sqrt{\cH_{R_k}(\rho)\cT_{R_k}(\rho)}\right.\\
  \label{3}
       & + \int_{B_k,\ \frac{p(x)}c<s}\rd
         x\left(\tfrac34\delta \rho(x)^\frac43
         -{\kappa\rho(x)\over|x-R_k|}\right)\\
  \label{4}
       &+ \int_{B_k,\ p(x)/c>s}\rd x
         \left(\tfrac38\delta
       \rho(x)^\frac43-\kappa H_k(x)\rho(x)\right)\\
  \label{5}
      &+\left.\int_{\Gamma_k\setminus
          B_k}\rd x\left(\tfrac34\delta\rho(x)^\frac43
        -{\kappa\rho(x)\over|x-R_k|}\right) \right]\\
  &+ \sum_{k=1}^K{Z^2\over8D_k}-C_cN.
\end{align}
We pick $s$ such that
\begin{equation}
  \label{s}
  2 \sqrt{{3^\frac53\lambda \over2^7\pi^\frac23}\frac38\delta}={\kappa
    \over\sqrt{\arsinh(s)}}
\end{equation}
which makes \eqref{2} a sum of complete squares. Next
\begin{equation}
  \label{3a}
  \begin{split}
    &\eqref{3}\geq \delta (cs)^4\inf\left\{\int_{p(x)<1}\rd
      x\left(\tfrac34 \rho(x)^\frac43
        -{\kappa\rho(x)\over cs\delta|x|}\right)\Big|\rho\in P\right\}\\
      \geq& {cs\kappa^3\over\delta^2}\inf\left\{\int_{p(x)<1}\rd
    x\left(\tfrac34 \rho(x)^\frac43 -{\rho(x)\over|x|}\right)\Big|\rho\in P\right\}\geq
        -C{cs\kappa^3\over\delta^2}
      \end{split} 
\end{equation}
where we replaced $p$ by $c sp$ in the first step and $x$ by
$\kappa/(cs\delta)x$ in the second step. Thus \eqref{3a} yields, after
summation, $K$ times a constant which is irrelevant for
stability. Furthermore,
\begin{equation}
  \label{4a}
  \eqref{4}\geq -{2^4\kappa^4\cdot4\pi\over2\cdot 4\delta^3D_k} \int_0^1\rd r r^22^4(1 +r^2/4)^2
  =-{5944\pi\kappa^4\over105\delta^3 D_k}
\end{equation}
and using $\int_{\Gamma_k\setminus B_k}\rd x|x-R_k|^{-4}\leq3\pi/D_k$
(Lieb et al. \cite[Formula (4.6)]{Liebetal1996}) we get
\begin{equation}
  \label{5a}
  \eqref{5}\geq -{ \kappa^4\over 4\delta^3}\int_{\Gamma_k\setminus
    B_k}{1\over|x-R_k|^4}\geq  -{3\pi \kappa^4\over4 \delta^3D_k}.
\end{equation}
Thus, the energy per particle is bounded from below uniformly in
$\rho$, $K$, and $N$, if
\begin{equation}
  \label{bedingung0}
  c\left({2972\pi\kappa^4\over 105\delta^3}+{3\pi\kappa^4\over4\delta^3}\right)\leq {Z^2\over8},
\end{equation}
i.e.,
\begin{equation}
  \label{bedingung}
  Z\leq Z_\mathrm{max}:=3{\sqrt{1686370\pi} \over 48182}c^\frac32.
\end{equation}
Numerically, using the physical value of the velocity of light $c=137.037$
(Bethe and Salpeter \cite[p. 84]{BetheSalpeter1957}), we get
\begin{equation}
  Z_\mathrm{max}\approx 229.9029615
\end{equation}
covering liberally all known elements. The result can be condensed into
\begin{satz}
  \label{satz2}
  There exists a constant
  $C$ such that for all $\rho\in P$ and all pairwise different
  $R_1,...,R_K\in\rz^3$ and $Z_1=...=Z_K\in [0,Z_\mathrm{max}]$
  $$\cE^\mathrm{TFWD}_{\mathfrak{R},\mathfrak{Z}}(\rho)\geq C\cdot(K+N).$$
\end{satz}
We conclude with two remarks:

1. Theorem \ref{satz2} holds actually for all
$\gZ\in[0,Z_\mathrm{max}]^K$. Our proof obviously generalizes to that
case, since the potential estimate \eqref{electrostat} also
generalizes in the obvious way.

2. It might be surprising that there is no requirement on a minimal
velocity of light which is independent of the value of $Z$. This is
different from other unrenormalized relativistic models like the
Thomas-Fermi functional with inhomogeneity correction
$(\sqrt\rho,|\nabla|\sqrt\rho)$ investigated by Lieb et
al. \cite[Formula (2.7)]{Liebetal1996}. There we were forced to
control the exchange energy by the Thomas-Fermi term. Here, this is no
longer necessary: due to the renormalization in Engel's and Dreizler's
derivation the exchange energy is bounded from below by a multiple of
the particle number.

\section*{Acknowledgments}
Special thanks go to Rupert Frank for many inspiring discussions, in
particular for pointing out that in addition to separation in high and
low density regimes, a localization in space near the nucleus could be
useful, and for critical reading of a substantial part of the
manuscript.

Thanks go also to Hongshuo Chen for critical reading of the manuscript.

Partial support by the Deutsche Forschungsgemeinschaft (DFG, German
Research Foundation) through Germany's Excellence Strategy EXC - 2111
- 390814868 is gratefully acknowledged.

%\bibliographystyle{plain}
%\bibliography{coulomb}

\begin{thebibliography}{10}

\bibitem{Benguria1979}
Rafael Benguria.
\newblock {\em The von {W}eizs{\"acker} and Exchange Corrections in the
  {T}homas-{F}ermi Theory}.
\newblock PhD thesis, Princeton, Department of Physics, June 1979.

\bibitem{Benguriaetal1981}
Rafael Benguria, Haim Brezis, and Elliott~H. Lieb.
\newblock The {T}homas-{F}ermi-von {W}eizs\"{a}cker theory of atoms and
  molecules.
\newblock {\em Comm. Math. Phys.}, 79(2):167--180, 1981.

\bibitem{BenguriaLieb1985}
Rafael Benguria and Elliott~H.\ Lieb.
\newblock The most negative ion in the {T}homas-{F}ermi-von {W}eizs{\"a}cker
  theory of atoms and molecules.
\newblock {\em J.\ Phys.\ B.}, 18:1045--1059, 1985.

\bibitem{Benguriaetal2008}
Rafael~D Benguria, Michael Loss, and Heinz Siedentop.
\newblock Stability of atoms and molecules in an ultrarelativistic
  {T}homas-{F}ermi-{W}eizs{\"a}cker model.
\newblock {\em Journal of Mathematical Physics}, 49(1):012302, 2008.

\bibitem{BetheSalpeter1957}
Hans~A.\ Bethe and Edwin~E.\ Salpeter.
\newblock Quantum mechanics of one- and two-electron atoms.
\newblock In S.\ Fl{\"u}gge, editor, {\em Handbuch der {P}hysik, {XXXV}}, pages
  88--436. Springer, Berlin, 1 edition, 1957.

\bibitem{CassanasSiedentop2006}
Roch Cassanas and Heinz Siedentop.
\newblock The ground-state energy of heavy atoms according to {B}rown and
  {R}avenhall: Absence of relativistic effects in leading order.
\newblock {\em J. Phys. A}, 39(33):10405--10414, 2006.

\bibitem{Chandrasekhar1931}
Subramanyan Chandrasekhar.
\newblock The maximum mass of ideal white dwarfs.
\newblock {\em Astrophys. J.}, 74:81--82, 1931.

\bibitem{Chenetal2020}
Hongshuo Chen, Rupert~L. Frank, and Heinz Siedentop.
\newblock A statistical theory of heavy atoms: Energy and excess charge.
\newblock arxiv:2010.12074, October 2020.

\bibitem{ChenSiedentop2020}
Hongshuo Chen and Heinz Siedentop.
\newblock On the excess charge of a relativistic statistical model of molecules
  with an inhomogeneity correction.
\newblock {\em Journal of Physics A: Mathematical and Theoretical},
  53(39):395201, September 2020.

\bibitem{FrankEkholm2006}
T.~Ekholm and R.~L. Frank.
\newblock On {L}ieb-{T}hirring inequalities for {S}chr\"{o}dinger operators
  with virtual level.
\newblock {\em Comm. Math. Phys.}, 264(3):725--740, 2006.

\bibitem{EngelDreizler1987}
E.~Engel and R.~M. Dreizler.
\newblock Field-theoretical approach to a relativistic
  {T}homas-{F}ermi-{D}irac-{W}eizs\"acker model.
\newblock {\em Phys. Rev. A}, 35:3607--3618, May 1987.

\bibitem{Fermi1927}
E.\ Fermi.
\newblock Un metodo statistico per la determinazione di alcune propriet{\'a}
  dell'atomo.
\newblock {\em Atti della Reale Accademia Nazionale dei Lincei, Rendiconti,
  Classe di Scienze Fisiche, Matematiche e Naturali}, 6(12):602--607, 1927.

\bibitem{Fermi1928}
E.~{Fermi}.
\newblock {Eine statistische Methode zur Bestimmung einiger Eigenschaften des
  Atoms und ihre Anwendung auf die Theorie des periodischen Systems der
  Elemente.}
\newblock {\em {Z. Phys.}}, 48:73--79, 1928.

\bibitem{Franketal2007S}
Rupert~L. Frank, Elliott~H. Lieb, and Robert Seiringer.
\newblock Stability of relativistic matter with magnetic fields for nuclear
  charges up to the critical value.
\newblock {\em Comm. Math. Phys.}, 275(2):479--489, 2007.

\bibitem{Franketal2008}
Rupert~L. Frank, Heinz Siedentop, and Simone Warzel.
\newblock The ground state energy of heavy atoms: Relativistic lowering of the
  leading energy correction.
\newblock {\em Comm. Math. Phys.}, 278(2):549--566, 2008.

\bibitem{Franketal2009}
Rupert~L. Frank, Heinz Siedentop, and Simone Warzel.
\newblock The energy of heavy atoms according to {B}rown and {R}avenhall: the
  {S}cott correction.
\newblock {\em Doc. Math.}, 14:463--516, 2009.

\bibitem{Gombas1949}
P.~Gomb{\'a}s.
\newblock {\em Die statistische {T}heorie des {A}toms und ihre {A}nwendungen}.
\newblock Springer-Verlag, Wien, 1 edition, 1949.

\bibitem{Gombas1956}
P.\ Gomb\'as.
\newblock {S}tatistische {B}ehandlung des {A}toms.
\newblock In S.\ {Fl\"ugge}, editor, {\em {H}andbuch der {P}ysik. Atome II},
  volume~36, pages 109--231. Springer-Verlag, Berlin, 1956.

\bibitem{HandrekSiedentop2013}
Michael Handrek and Heinz Siedentop.
\newblock On the maximal excess charge of the {C}handrasekhar-{C}oulomb
  {H}amiltonian in two dimension.
\newblock {\em Lett. Math. Phys.}, 103(8):843--849, 2013.

\bibitem{Lieb1982A}
Elliott~H. Lieb.
\newblock Analysis of the {T}homas-{F}ermi-von {W}eizs\"{a}cker equation for an
  infinite atom without electron repulsion.
\newblock {\em Comm. Math. Phys.}, 85(1):15--25, 1982.

\bibitem{LiebLiberman1982}
Elliott~H. Lieb and David~A. Liberman.
\newblock Numerical calculation of the {T}homas-{F}ermi-von {W}eizs\"acker
  function for an infinite atom without electron repulsion.
\newblock Technical Report LA-9186-MS, Los Alamos National Laboratory, Los
  Alamos, New Mexico, April 1982.

\bibitem{Liebetal1996}
Elliott~H.\ Lieb, Michael Loss, and Heinz Siedentop.
\newblock Stability of relativistic matter via {T}homas-{F}ermi theory.
\newblock {\em Helv.\ Phys.\ Acta}, 69(5/6):974--984, December 1996.

\bibitem{LiebSimon1977}
Elliott~H. Lieb and Barry Simon.
\newblock The {T}homas-{F}ermi theory of atoms, molecules and solids.
\newblock {\em Advances in Math.}, 23(1):22--116, 1977.

\bibitem{LiebYau1988}
Elliott~H.\ Lieb and Horng-Tzer Yau.
\newblock The stability and instability of relativistic matter.
\newblock {\em Comm.\ Math.\ Phys.}, 118:177--213, 1988.

\bibitem{MarchYoung1958}
N.~H.\ March and W.~H.\ Young.
\newblock Variational methods based on the density matrix.
\newblock {\em Proc.\ Phys.\ Soc.}, 72:182--192, 1958.

\bibitem{Sorensen2005}
Thomas {\O}stergaard~S{\o}rensen.
\newblock The large-{$Z$} behavior of pseudorelativistic atoms.
\newblock {\em J. Math. Phys.}, 46(5):052307, 24, 2005.

\bibitem{Simon1979}
Barry Simon.
\newblock {\em Functional Integration and Quantum Physics}.
\newblock Academic Press Inc. [Harcourt Brace Jovanovich Publishers], New York,
  1979.

\bibitem{Solovejetal2008}
Jan~Philip Solovej, Thomas {\O stergaard S{\o}rensen}, and Wolfgang~L. Spitzer.
\newblock The relativistic {S}cott correction for atoms and molecules.
\newblock {\em Commun. Pure Appl. Math.}, 63:39--118, January 2010.

\bibitem{Thomas1927}
L.~H.\ Thomas.
\newblock The calculation of atomic fields.
\newblock {\em Proc.\ Camb.\ Phil.\ Soc.}, 23:542--548, 1927.

\bibitem{Tomishima1969}
Yasuo Tomishima.
\newblock {A Relativistic Thomas-Fermi Theory}.
\newblock {\em Progress of Theoretical Physics}, 42(3):437--447, 09 1969.

\bibitem{Weizsacker1935}
C.~F. v.~Weizs{\"a}cker.
\newblock Zur {T}heorie der {K}ernmassen.
\newblock {\em Z.\ Phys.}, 96:431--458, 1935.

\bibitem{YoneiTomishima1965}
Katsumi Yonei and Yasuo Tomishima.
\newblock On the {W}eizs{\"a}cker correction to the {T}homas-{F}ermi theory of
  the atom.
\newblock {\em Journal of the Physical Society of Japan}, 20(6):1051--1057,
  1965.

\end{thebibliography}

\def\cprime{$'$}

\end{document}